\documentclass[singlespacing]{elsart}

\usepackage{graphicx}

\usepackage{amssymb}
\journal{}
\begin{document}

\begin{frontmatter}

\title{Dynamical contribution into enzyme catalytic efficiency}

\author{A.E. Sitnitsky},
\ead{sitnitsky@mail.knc.ru}

\address{Institute of Biochemistry and Biophysics, P.O.B. 30, Kazan
420111, Russia}

\begin{abstract}
A realistic physical model for the so called rate promoting
vibration (RPV) at enzyme action  is constructed. The origin of
the RPV is assumed to be an oscillating electric field produced
by long-lived localized vibrational modes in protein dynamics,
namely, by the so called discrete breather (DB) in secondary
structure. The strength of interaction of the RPV with the
reaction coordinate is evaluated and its effect on the reaction
acceleration is assessed within the framework of modern
theory for thermally activated escape rate at periodic driving. We
reveal the phenomenon of resonant activation in our model
elucidating why the frequency of the RPV in the range
$100\div200\  cm^{-1}$ was chosen by the evolution of enzymes as
an optimal one. The effect of the RPV on the reaction
acceleration is shown to vary from moderate one (up to
$10^3\div10^4$) in the case of three-site DB to enormous (up to
$10^6\div10^8$) in the case of five-site DB and thus can
significantly contribute into enzyme catalytic efficiency. Also
the model is shown to be compatible with the known functional
dependence of
enzymatic reaction rates on  solvent viscosity.

\\

\end{abstract}

\end{frontmatter}

\section{Introduction}

Comprehension that specific protein dynamics contributes somehow
into enormous reaction acceleration (up to $10^8\div10^{15}$) by
enzymes has a long history (see \cite{Kar79}, \cite{Wel82},
\cite{Kar83}, \cite{War84}, \cite{Flu85}, \cite{Ner97},
\cite{Alp01}, \cite{Agr05} and refs. therein). It is supposed
that an enzyme provides more efficient reaction activation
compared to corresponding nonenzymatic reaction due to a
dynamical mechanism and that the latter is significant enough to
be a rival to the effect of lowering of the potential energy
barrier (due to transition state stabilization) caused by
specific interactions of the substrate molecule in an enzyme
active site with catalytically active groups. The latter paradigm
going back to Pauling is a cornerstone of traditional chemical
enzymology \cite{Fer99} and is developed in a number of
vigorously conquering approaches such as transition state
stabilization by: i) electrostatic field of preorganized dipoles
in an enzyme active site \cite{War97}, \cite{War98}, ii) low
energy barrier hydrogen bonds \cite{Cle98}, iii) entropic, strain
and desolvation effects (leading mostly to ground state
destabilization) \cite{Fer99}, etc. Chemical enzymology is a vast
field of activity and it will not be touched upon here. Instead
we deal with physical aspects of enzyme action which are still
poorly understood. The reason for that seems to be in the lack of
adequate theoretical tools to evaluate the contribution of a
dynamical mechanism in the reaction rate and in meager direct
experimental evidence for its existence. The latter is altered to
some extent by recent experiments \cite{Ant01}, \cite{Sut02},
\cite{Ant02}, \cite{Kna02}, \cite{Min03}, \cite{Min04}. In these
papers the importance of protein dynamics for hydrogen tunneling
transfer enzymatic reactions was shown and the notion of the rate
promoting vibration (RPV) was coined \cite{Ant01}, \cite{Ant02}
(some authors call it protein promoting vibration). To regret
hydrogen tunneling transfer is involved only in minor and
marginal part of enzymatic reactions and up to now there is no
direct experimental evidence for the significance of the RPV in
covalent bond cleavage between heavy atoms that is the most
typical rate limiting step at enzyme action. For such enzymes the
well known solvent viscosity effect on the reaction rate (see
\cite{Flu85}, \cite{Yed95} and refs. therein) still remains the
most vivid testimony for manifesting itself of protein dynamics
in enzyme catalysis. It is commonly accepted that this phenomenon
is mediated by protein dynamics but the detailed mechanism is
obscure.

An important question is what kind of motion in proteins is
responsible for the dynamical contribution into enzyme catalytic
efficiency? Some authors relate protein specific motion producing
the RPV with conformational motion of peculiar amino-acid residues
which may belong to enzyme active cite or be at a distance from it
\cite{Min03}, \cite{Min04}. In our opinion this scenario
encounters difficulties at attempts to reconcile it with the
experiments \cite{Dan95}, \cite{Dan98}, \cite{Bra00},
\cite{Dun00}. The latter unequivocally testify that enzymes retain
their catalytic capacity at temperatures below the so-called
dynamical "glass" transition where conformational motion is
considerably suppressed (see, e.g., \cite{Ibe89}, \cite{Dan98}
and refs. therein). Some guess on the origin of the RPV can be
obtained by considering its frequency. The authors of
\cite{Ant02} conclude that the estimated dominant peaks in the
spectral densities of the RPV indicate motions on the $150\
cm^{-1}$ frequency scale. There are another assessments of this
value in literature: $\le 300\ cm^{-1}$ \cite{Ner97} and $200\
cm^{-1}$ \cite{Kov99}. In the latter paper the RPV is attributed
to the mode of Amide-VII of a peptide group vibrations. However
this mode requires some torsion of the plane of a peptide group
and is very energetically consumptive and unfavorable. In our
paper \cite{Sit03} it was shown that there is another possible
type of long-lived localized vibrational modes in protein
dynamics with required frequency, namely, the so called discrete
breather (DB) in protein secondary structures. The non-linear
localized excitations named DBs or else intrinsic localized modes
are time periodic spatially localized oscillations with
significant amplitudes of several units in a chain of weakly
coupled non-linear oscillators while others are at rest or
oscillate with negligible amplitudes. They were discovered by
Sievers and Takeno in 1988 \cite{Siev88} and proved by MacKay and
Aubry \cite{Mac94} to be structurally stable. By now they are
well understood and commonly appreciated as a generic phenomenon
in nature (see \cite{Fla98} and refs. therein). DBs have become a
new and very fruitful paradigm in nonlinear physics and in
particular much work is being carried out at present to construct
models for DBs in biomolecules. The paper \cite{Sit03} is a
development along this line. It is shown there that the frequency
of a DB in an $\alpha$-helix can be obtained equal to $115\ cm^{-1}$
in accordance with the results of the experiments of
\cite{Xie02} on far infra-red laser pulse
spectroscopy of proteins. The DB is actually an oscillation of
the planes of some peptide groups in an $\alpha$-helix or
$\beta$-sheet around their equilibrium positions with considerable
amplitude (up to $10^{\circ}\div15^{\circ}$) while neighboring
peptide groups oscillate with much less amplitudes and more
distant groups oscillate with negligible ones or stay at rest.
Thus the DB is assumed to store and utilize the energy released
at substrate binding by an enzyme. It is created by an external
cause (binding energy) rather than by equilibrium thermal
fluctuations though can be fed by the latter. As a peptide group
is known to have a large dipole moment ($\approx 3.6\ D$)
parallel to its plane \cite{Can80} the participation of the
latter in the DB creates an oscillating electric field. Such
fields (the so called electrostatic fluctuations) are for long
supposed to be a key factor for enzyme catalysis \cite{War84}.
The field can interact with the reaction coordinate because
during the movement of the system along the latter a separation
of charges takes place yielding the dipole moment of the reaction
coordinate. Thus protein dynamics affects the potential surface
for the reaction making the latter to be non-stationary. The
shape (e.g., the height) of the reaction energy barrier becomes
time dependent. As will be discussed below the oscillating
electric field in accordance with the general theory of the
effect of periodic driving force \cite{Dyk01}
""heats up" the system by changing
its effective temperature thus giving rise to lowering of the
activation energy of escape which can be much bigger than the
real temperature even for comparatively weak fields".

The aim of the present paper is to construct a realistic physical
model for the dynamical mechanism contributing into enzyme action
and to evaluate its possible catalytic efficiency. We attain the
latter within the framework of the archetypical model for the
reaction rate, namely that of the overdamped limit
of the Fokker-Planck equation (see \cite{Han90} and refs
therein) with non-stationary force field. This equation is a
standard tool for studying phenomena with time dependent
potentials mostly within the context of stochastic resonance
\cite{Jun93}, \cite{Gam98}, thermally activated escape problem at
periodic driving in the classical regime \cite{Leh00},
\cite{Leh000}, \cite{Leh03}, \cite{Sme99}, \cite{Sme99a},
\cite{Dyk01}, \cite{Mai01}, \cite{Tal99}, \cite{Tal04}, \cite{Ryv04},
\cite{Dyk05}, \cite{Dyk051} and resonant activation
(see \cite{Rei95} and refs therein).
For the estimate of the escape rate we apply the results of the
papers \cite{Leh000}, \cite{Sme99}, \cite{Dyk051}.

The paper is organized as follows. In Sec.2 we consider the
 oscillating electric field
created by a DB in protein secondary structure. In Sec.3 we
evaluate the ability of this field to accelerate an enzymatic
reaction. In Sec.4 we argue that our model is able to shed new
light on the origin of solvent viscosity effect on enzymatic
reaction rates. In Sec.5 the results are discussed and the
conclusions are summarized.

\section{Physical model for the rate promoting vibration}

The possibility for existence of a stable zig-zag shape DB in
protein secondary structures ($\alpha $-helix or $\beta $-sheet)
was shown in the paper \cite{Sit03}. The DB in our case is
actually the out of phase oscillations of the planes of some
adjacent peptide groups (due to rotations round their torsional
angles) with considerable amplitudes up to $\approx10^{\circ}\div
15^{\circ}$ from their equilibrium values while neighboring
peptide groups oscillate with much less amplitudes and more
distant groups oscillate with negligible ones or stay at rest.
The dimensionless form of the equation of motion for the small
deviation $x_i(t)$ ($x\leq0.3$) of the angle of the plane for the
i-th peptide group from its equilibrium value in a chain
$i=1,...,N$ is
\begin{equation}
\label{eq1} \frac{d^2x_i}{dt^2}=\rho (x_{i-1}+x_{i+1})-U'(x_i)
\end{equation}
where $\rho$ is the coupling constant, $U(x_i)$ is a nonlinear
local potential and the dash denotes a derivative in coordinate.
 In \cite{Sit03} the following results were
obtained:

$\alpha $-helix:\ \ $ \rho=-0.00193;\ \ U(x)\approx0.5 x^2 +
0.017 x^3 +...$.

anti-parallel $\beta $-sheet:\ \ $ \rho=-0.00115;\ \
U(x)\approx0.5 x^2 + 0.004 x^3 +...$.

parallel $\beta $-sheet:\ \ $ \rho=0.0028;\ \  U(x)\approx0.5 x^2
- 0.3 x^3 + ...$.

Since parallel $\beta $-sheet is practically not represented in
natural proteins \cite{Bra99} we omit it from further
consideration. The linearizeation of (\ref{eq1}) yields the
dispersion relationship
$\omega_k(\rho)=\sqrt{1+2\left|\rho\right|cos (k)}$ where $k$ is
the wave number. The DB frequency and its harmonics must not coincide with linear
modes spectrum \cite{Fla98}, i.e., satisfy the requirement
$\omega_{DB}
> \omega_k (\rho)$. There is no analytical solution of (\ref{eq1}) for the general
case of a multi-site DB \cite{Fla98}. However numerical
simulations show that in our case of the DB in protein secondary
structure the time dependence $x_i(t)$ is qualitatively rather
similar to harmonic oscillations \cite{Sit03}. The latter can be
easily anticipated from (\ref{eq1}) taking into account the
smallness of the coupling constants $\rho$ and of the anharmonic
corrections in the local potentials. That is why we restrict
ourselves by a harmonic approximation as sufficient for further
analysis
\begin{equation}
\label{eq2} x_i(t)=x_i^m f(\omega_{DB}t+\phi_i); \ \ \ \ \
f(t)=sin(t)
\end{equation}
where $x_i^m$ is the amplitude and $\phi_i$ is the phase. At first
we consider the simplest case of a three-site DB. In this case we
can take into account only the motion of the central peptide group
in the three-site DB undergoing maximal deviations and neglect
others because in such DB the amplitude of the central peptide
group by an order of magnitude exceeds those of the neighbors
$x^m \equiv x_i^m >> x_{i\pm1}^m$ \cite{Sit03}. Returning to
dimensional time we should replace $\omega_{DB}$ by dimensional
 frequency $\Omega$
($\Omega=\omega_{DB}\sqrt{\frac{V"_{eff}(0)}{I}}$) where $V_{eff}$
is the effective local potential and $I$ is the moment of inertia
of the peptide group. In \cite{Sit03} the following results for
$\sqrt{\frac{V^{\prime\prime}_{eff}(0)}{I}}$ were obtained :\ 114.75 $cm^{-1}$
for $\alpha $-helix; 84.34 $cm^{-1}$ for anti-parallel $\beta
$-sheet and 85.49 $cm^{-1}$ for parallel $\beta $-sheet, i.e.,
approximately $2\times 10^{13}s^{-1}$ for all of them. Now let us
take into account that the peptide group is known to have a large
dipole moment ($p\approx 3.6\ D$) parallel to its plane
\cite{Can80}. In an $\alpha$-helix the dipole moment of a peptide
group is approximately parallel to its direction. If we place the
origin of the Cartesian frame in the dipole moment and direct the
axis X along the dipole moment at equilibrium position of the
peptide group (and consequently along the axis of the $\alpha
$-helix) then the dipole moment at time t has the projections:
$p_i^X(t)=p\ cos(x^m f(\Omega t))$,\ $p_i^Y(t)=p\ sin(x^m
f(\Omega t))$,\ $p_i^Z(t)\approx 0)$ (see Fig.1). Taking into
account that $x^m f(\Omega t)<<1$ we have: $\bar p_i(t)\approx p\
\bar e_X+p\ x^m f(\Omega t)\ \bar e_Y$.

We are going to examine the influence of the electric field
produced by this oscillating dipole moment on the enzymatic
reaction, i.e., on the covalent bond breaking between heavy atoms
that is the most typical rate limiting step at enzyme action. We
assume that the reaction coordinate $q$ is located at a distance
$R_i$ from the dipole moment in the point $\bar R_i=R_i\ \bar e_Y$ and
its direction constitutes the angle $\alpha$ with the axis $Y$,
i.e., $\left(\bar e_q\cdot\bar e_Y\right)=cos\ \alpha$ (see
Fig.1). The quasistationary electric field ($\Omega<<c/R_i$ where
$c$ is the speed of light) from the dipole moment in this point is
\begin{equation}
\label{eq3} \bar E(t)=\frac{3\left(\bar p_i(t)\cdot\bar R_i\right)\bar
R_i-R_i^2\bar p_i}{\varepsilon R_i^5}=\frac{p}{\varepsilon R_i^3}\ \bar
e_X+\frac{2\ p\ x^m f(\Omega t)}{\varepsilon R_i^3}\ \bar e_Y
\end{equation}
where $\varepsilon$ is the dielectric constant for protein
interior ($\varepsilon\approx3.5$\  \cite{Can80}). We attribute
the dipole moment $\bar d=e\ q\ \bar e_q$ to the bond to be
broken at the reaction where $e$ is the partial charge. The
energy of the interaction of the electric field with the dipole
moment $\bar d$ is
\begin{equation}
\label{eq4} H_{int}(t)=-\left(\bar d\cdot\bar
E(t)\right)=const-\frac{2\ e\ p\ x^m\ cos\ \alpha}{\varepsilon
R_i^3}\ q\ f(\Omega t).
\end{equation}
The factor before $qf(\Omega t)$ in the right hand
side is the strength of the interaction.

Now we can qualitatively consider the case of a multi-site DB
where several peptide groups oscillate with non-negligible
amplitudes. For instance in a five-site DB three of them have
significant values compared with the neighbors ($x_i^m \simeq
x_{i\pm1}^m >> x_{i\pm2}^m$). This example is quite
representative and we further restrict ourselves by it. On the one hand
peptide groups in the zig-zag DB undergo out of phase
oscillations, i.e., the deviations of the angular variables have
opposite signs ($sgn \bigl( \delta x_i (t) \bigr)= - sgn \bigl( \delta
x_{i\pm1} (t) \bigr)$). On the other hand the structure of an
$\alpha$-helix is such that there are three peptide groups in its
perpendicular cros-section, i.e., the radial angle between
adjacent groups is $120^{\circ} > 90^{\circ}$
(actually there are 3.6 peptide groups in the
perpendicular cros-section, so that the angle is
$\approx 100^{\circ}$ \cite{Can80} but this subtliety does not modify our
 futher qualitative conclusions). Two
circumstances compensate each other. That is why at opposite
deviations of the angular variables the projections of linear
deviations of the dipole moments of adjacent peptide groups  on
the axis $Y$ have the same sign, i.e., are in phase (see
Fig.1). If we neglect for simplicity the differences both
in the distances from the dipoles to the location of the reaction
coordinate and in angles between the dipoles and the direction of the
reaction coordinate then we can assert that the oscillating
electric field produced by three peptide groups of the five-site
DB has the amplitude that is approximately thrice of that for
one peptide group of the three-site DB and consequently the strength of
the interaction in (\ref{eq4}) should be multiplied by the factor
$3$. Taking into account the above mentioned differences makes it
of course less than $3$ and we use the qualitative factor $2\div3$ as
sufficient for our further analysis.

In the next Sec. we consider the effect of the time dependent part
of the interaction (\ref{eq4}) on the enzymatic reaction.

\section{Efficiency of the rate promoting vibration
for reaction acceleration}

We use the Kramers' model for a chemical reaction as the escape
of a Brownian particle with the mass $m$ from the well of a
metastable potential (see for review \cite{Han90} and
refs. therein) in the presence of the additional time dependent
term defined by (\ref{eq4}). For the metastable potential we choose
without any serious loss of generality the conventionally
accepted simple form $V(q)=aq^2/2-bq^3/3$. This potential has
a minimum at $q_{min}=0$, a maximum at $q_{max}=a/b$ and the
barrier height
$V(q_{max})-V(q_{min})=a^3/(6b^2)$. In the overdamped limit (high
friction case) the Langevin equation of motion along the reaction
coordinate $q$ is
\begin{equation}
\label{eq5} \gamma
\frac{dq}{dt}=-\frac{aq}{m}+\frac{bq^2}{m}+\frac{2\ e\ p\ x^m\ cos\
\alpha}{m\ \varepsilon R^3}\ f(\Omega t)+\sqrt{2\gamma k_B T/m}\
\xi(t)
\end{equation}
where $\gamma$ is the friction coefficient, $k_B$ is the Boltzman
constant, $T$ is the temperature and $\xi(t)$ is the Gaussian
white noise ($<\xi(t)\xi(0)> =\delta (t)$ where $\delta (t)$ is
the Dirac $\delta$-function) with zero mean ($<\xi(t)>=0$). We
introduce the dimensionless variables and parameters
\begin{equation}
\label{eq6} \tilde q=q b/a;\ \ \ \ \  \tilde
t=\frac{ta}{\gamma m};\ \ \ \ \  \tilde \Omega=\frac{\Omega \gamma
m}{a}
\end{equation}
and denote
\begin{equation}
\label{eq7} A=\frac{2\ e\ p\ x^m\ cos\ \alpha}{a\ \varepsilon
R^3}\ b/a ;\ \ \ \ \ D=\frac{k_B T\ b^2}{a^3}=\frac{k_B
T}{6\left[V(q_{max})-V(q_{min})\right]}
\end{equation}
where $A$ is the dimensionless parameter characterizing the
oscillating electric field strength and $D$ is the dimensionless
parameter characterizing the intensity of thermal fluctuations
(i.e., the temperature of the heat bath) relative the barrier
height. Their ratio is the most important parameter of the model
\begin{equation}
\label{eq8} \frac{A}{D}=\frac{2\ e\ p\ x^m\ cos\
\alpha}{\varepsilon R^3k_B T}a/b.
\end{equation}
Omitting the overlines we obtain
\begin{equation}
\label{eq9} \frac{dq}{dt}=q-q^3+Af(\Omega t)+\sqrt{2\gamma Dm/a}\
\xi(t\gamma m/a).
\end{equation}
The corresponding overdamped limit of the  Fokker-Planck
equation for the probability
distribution function $P(q,t)$ \cite{Han90} is
\begin{equation}
\label{eq10} \frac{\partial P(q,t)}{\partial t}=-\frac{\partial
}{\partial q}\Bigl\lbrace [-q+q^2+Af(\Omega
t)]P(q,t)\Bigr\rbrace+D\frac{\partial^2P(q,t)}{\partial q^2}.
\end{equation}

Before proceed further we should define the range of the
parameters for our model. To be supported by evidence we take the
numerals for a very typical and one of the most studied enzymatic
reaction catalysed by Subtilisin. This enzyme belongs to serine
proteases and brakes the bond between the atoms C and N in a
peptide group of a substrate of protein nature. At physiological
temperatures where enzymes normally work ($k_B T \approx
4.2\cdot10^{-14}\ CGS$) the corresponding noncatalysed reaction
typically has the rate constant $k\approx 1 \cdot
10^{-8}\  s^{-1}$ while the enzymatic reaction has the
rate constant (that of the rate limiting step) $k_{cat}\approx 5
\cdot 10^1\  s^{-1}$ \cite{Car88}. Thus the total catalytic
effect is $\approx 10^{12}$. Usually we do not know the friction
coefficient $\gamma$. However assuming the single local-system
relaxation time $\gamma m/a$ \cite{Han90} to be typically $\leq
0.5\div 1.5\cdot 10^{-14} s$ we conclude that for the noncatalysed reaction we
must have $\Gamma=\frac{\gamma m k}{a}\approx 10^{-22}$. We can
evaluate from the Arrhenius factor of the Kramers' rate
\begin{equation}
\label{eq11} \Gamma_K =(\sqrt 2 \pi)^{-1} exp\biggl[-1/(6D)\biggr]
\end{equation}
that in order to obtain such a reaction rate constant in the
absence of the oscillating electric field ($A=0$) we should take
$D\approx \frac{1}{6\cdot 22\  ln 10} \approx 3.3\cdot 10^{-3}$. We also assume
the distance from the minimum to the barrier to be typically
$a/b\approx0.5\ \AA$. Taking the typical values (
$e\approx0.5$ of the electron charge ($4.8\cdot10^{-10}\ CGS);\
p=3.6D=3.6\cdot10^{-18}\ CGS;\ x_m\leq0.3;\ \varepsilon\approx
1\div 3.5$) and assuming that the DB is arranged favorably
respective the reaction coordinate ($\alpha\approx0$) we obtain
that for $R\geq2 \AA$ the ratio is $A/D\leq10$ for the case of
the three-site DB and can attain the values $A/D\leq20\div30$ for
the five-site DB. It should be stressed that even in the latter
case $A\leq 0.1$ at our value of $D\approx 3.3\cdot 10^{-3}$ and
satisfies the requirement $\vert A \vert <0.25$,
i.e., we are in the so called subthreshold driving regime.
The latter provides that the potential surface
always has a minimum and a maximum, i.e., the oscillating field is small
enough not to distort the physical picture of the chemical
reaction as the Brownian particle escape from the metastable
state. We obtain from the dimensional frequency $\Omega\approx
2\times 10^{13}s^{-1}$ that the dimensionless one (see
(\ref{eq6})) is $\Gamma_K << \Omega\leq 0.1\div0.3$. It should be
stressed that we are very far from the stochastic resonance
regime defined by the requirement $\Omega=\pi \Gamma_K$
\cite{Gam98} and are in the so called semi-adiabatic regime
 \cite{Tal99}. Thus we define the range for all three parameters
figuring in (\ref{eq10})
\[
D\approx 3.3\cdot 10^{-3};\ \ \ \ \ \ \Omega\leq 0.1\div0.3;\ \ \ \ \ \ A/D\leq10
\]
for the three-site DB and
\begin{equation}
 \label{eq12}  D\approx 3.3\cdot 10^{-3};\ \ \ \ \ \ \Omega\leq 0.1\div0.3;\ \ \ \ \ \ A/D\leq20\div30
\end{equation}
for the five-site DB. The range of values for noise intensity $D$ in (\ref{eq12}) is
inaccessible for numerical solution of (\ref{eq10}) because of
enormous escape times ($1/\Gamma_K \geq 10^{22}$).
 That is why we resort to the
analytic results of the papers \cite{Leh000}, \cite{Sme99}, \cite{Dyk051}
 for the escape rate $\overline\Gamma$ averaged over the period of
the driving force oscillations $2\pi/\Omega$. We compile the results of
these papers
in a simple phenomenological formula for the
prefactor $F$ of the escape rate in the presence of the
harmonic driving force with the  intensity $A$ and
the frequency $\Omega$. For our potential
($V(q)=q^2/2-q^3/3$ in the dimensionless form)
 and in the semi-adiabatic
regime $ \Gamma_K << \Omega <1$  it takes the form
\[
 F\equiv
\frac{\overline\Gamma}{\Gamma_K}\approx\frac{\Omega}{2\pi}
\int\limits_{0}^{2\pi}d\phi
\]
\begin{equation}
\label{eq13}
exp\Biggl[-\frac{A}{D}\Biggl(\Biggl(g(\Omega)sin(\phi)\Biggr)+
\frac{A}{8D\Omega} \Biggl(\phi-\frac{sin(2\phi)}{2}\Biggr)
\Biggr)\Biggr]
\end{equation}
where $\Gamma_K$ is the Kramers' escape rate (\ref{eq11}) and
$g(\Omega)=\frac{\pi \Omega}{sh(\pi \Omega)}$.
 For $A/D >> 1$ the integral in
(\ref{eq13}) can be evaluated by the steepest descent method and
yields
\begin{equation}
 \label{eq14} F\approx\frac{\sqrt D}
 {\sqrt {2\pi A g(\Omega)}}\
 exp\Biggl[\frac{A}{D}\biggl(g(\Omega)-
\frac{A 3 \pi}{16\Omega}\biggr)\Biggr].
\end{equation}
The dependencies of the prefactor $F$ on $A/D$ and $\Omega$ are depicted
in Fig.2 and Fig.3 respectively. They are obtained from (\ref{eq13}) by
numerical integration. For the potential $V(q)=-q^2/2+q^4/4$ the
function $g(\Omega )$ is $g(\Omega)=\sqrt {\frac{\pi \Omega}{sh(\pi
\Omega)}}$.
We further use (\ref{eq14}) for the estimate of the catalytic
efficiency for the dynamical
contribution (that is actually the prefactor) at enzyme action.

\section{Solvent viscosity effect on the rate of enzymatic reaction}
It is well known that enzymatic reaction rates exhibit unusual
dependence on solvent viscosity $\eta$ \cite{Yed95}
\begin{equation}
 \label{eq15} k_{cat}\propto (\eta/\eta_0)^{-\beta(\sigma)}
\end{equation}
where $\eta_0=1\  cP$, $0\leq\beta\leq1$ that is generally of
order of $0.4\div0.6$ and $\sigma$ is the cosolvent molecular
weight. This dependence is the more pronounced the less is the
cosolvent molecular weight, i.e., the higher its capacity to
penetrate in protein interior \cite{Yed95}. The empirical law
(\ref{eq15}) is usualy verified in the range
 $1\geq\eta/\eta_0 < 50 \div 100$. As is commonly
accepted the mechanism of this dependence should be somehow
mediated by protein
dynamics but its origin is still not clear.

Below we argue that the present approach can shed some light on
the  empirical relationship (\ref{eq15}). We recall that for both
$\alpha$-helix and anti-parallel $\beta $-sheet the local
potentials for the DB are of hard
type \cite{Sit03} for which the frequency decreases with the
decrease of the amplitude of oscillations. As a result we come to
the following scenario. The increase of the viscosity leads to the
increase of the probability for a cosolvent molecule to be
located near the oscillating peptide group of the DB that
restricts the amplitude of the oscillations simply by creating
steric hindrance. The more the number of cosolvent molecules in
the vicinity of the DB the less the amplitude of the
oscillations. As a result of decreasing amplitude the frequency
of the DB is decreased. Now we phenomenologically find the law for
this decrease to obtain the required scaling of the reaction rate
constant  with the solvent viscosity (\ref{eq15}).

 In order to obtain from (\ref{eq14}) the required
empirical law (\ref{eq15}) we should have
\begin{equation}
 \label{eq16} \frac{exp\biggl[\frac{A}{D}g(\Omega)\biggr]}
 {\sqrt {\frac{ A g(\Omega)}{D}}}=\frac{C}
{\Bigl(\frac{\eta}{\eta_0}
 \Bigr)^{\beta (\sigma)}}
\end{equation}
where $C$ is a constant. We take into account that $A\propto x^m$ and that
frequency of the DB is some function (that is not kwown in the analytical form)
of the amplitude of oscillations $\Omega =\Omega(x^m)$. In pure
solvent without cosolvent
 molecules ($\eta=\eta_0$) we have some amplitude of the DB $x^m_0$ such
that the frequency mathes the requirement of the resonant activation
 $\Omega(x^m_0)\approx 0.2 \div 0.3$. Then we can determine the constant $C$
and write
\begin{equation}
 \label{eq17}  \frac{exp\biggl[\frac{A(x^m_0)}{D}\frac{x^m(\eta)}{x^m_0}
g\Bigl(\Omega\bigl(x^m(\eta)\bigr)\Bigr)\biggr]}
 {\sqrt{x^m(\eta)g\Bigl(\Omega\bigl(x^m(\eta)\bigr)\Bigr)}}=\frac{1}
{\Bigl(\frac{\eta}{\eta_0}
 \Bigr)^{\beta (\sigma)}}\frac{exp\biggl[\frac{A(x^m_0)}{D}
g\Bigl(\Omega(x^m_0)\Bigr)\biggr]}
 {\sqrt {x^m_0g\Bigl(\Omega(x^m_0)\Bigr)}}.
\end{equation}
This relationship implicitly defines the decrease of the amplitude of
oscillations $x^m(\eta)$ with the increase of solvent viscosity $\eta$.
Let us demonstrate it explicitly. We suppose that we are in the the region
$\eta/\eta_0 >> 1$ where $\Omega(x^m(\eta)) < \Omega(x^m_0)$ so that
 $g\Bigl(\Omega\bigl(x^m(\eta)\bigr)\Bigr)\rightarrow 1$. If we denote
 $\varepsilon=x^m(\eta)/x^m_0$ then we obtain from (\ref{eq17})
\begin{equation}
 \label{eq18} \varepsilon=\Bigl(\frac{\eta}{\eta_0}
 \Bigr)^{2\beta (\sigma)}g\Bigl(\Omega\bigl(x^m_0\bigr)\Bigr)
exp\Biggl[-\frac{2A(x^m_0)}{D}
\Biggl(g\Bigl(\Omega\bigl(x^m_0\bigr)\Bigr)
-\varepsilon \Biggr) \Biggr].
\end{equation}
Taking into account that most probably $\frac{A(x^m_0)}{D}\approx 10\div15 $
(see next Sec.) we see that the solution of this equation is
 $\varepsilon \rightarrow
0$, i.e., indeed $x^m(\eta)$ is sharply decreased copared with $x^m_0$ at
large $\eta/\eta_0$.
The construction of a physical model
 for solvent viscosity effect
on the amplitude of the DB $x^m(\eta)$ and for the dependence of the
 frequency of the DB on the amplitude of oscillations $\Omega(x^m)$
is beyond the scope of the present paper and will be
the subject of a separate investigation.

Concluding this Sec. the following comment should be done. Let us write the
full expression for the logarithm of the escape rate obtained from
(\ref{eq14}) and (\ref{eq11})
\[
 ln\overline\Gamma\approx
\]
\begin{equation}
 \label{eq19}  const-\frac{1}{6D}+
\frac{A}{D}\biggl(g(\Omega)-
\frac{A 3 \pi}{16\Omega}\biggr)+ln\biggl(\sqrt{D}\biggr)-
ln\biggl(\sqrt{2\pi A g(\Omega)}\biggr).
\end{equation}
This formula predicts two effects that at first glance can be
detected in the experiment. First the factor
$ln\biggl(\sqrt{D}\biggr)$ leads to a deviation of the Arrhenius
plot from the straight line. However the deviation is negligible
because this factor is fully overshadowed by a much more
significant dependence $\frac{1}{6D}$. Second the factor
$\frac{A}{D} g(\Omega)$ suggests that not only the prefactor, but
also the exponent in the escape rate should scale with the
viscosity of the solvent due to the dependence of $\Omega$ on it.
However this dependence also seems to be hardly observable
because the driving amplitude is very small $A << 1/6$. In the
next Sec. it will be argued that at present the traditional
chemical enzymology leaves no reasons to expect the driving
amplitude to noise intensity ratio to be very large. Most
probably we have $A/D \leq15$, i.e., at our value
 $D\approx 3.3\cdot 10^{-3}$ we obtain $A\leq0.06$. To regret this
 value seems
to be too small compared with $1/6 \approx 0.17$ for the
 dependence of the exponent on solvent viscosity to be noticed in the
experiment. No matter how insignificant the effect is, this specific
 prediction
remains the most direct way to verify the theory experimentally.

\section{Discussion}

We construct a realistic physical model in which the origin of the
RPV at enzyme action is the oscillating electric field produced by
dipole moments of peptide groups participating in long-lived
localized vibrational modes (i.e., in a DB) of protein dynamics.
We evaluate the magnitude of the oscillating electric field from
peptide groups in two cases. In the first case the field is
produced by a single peptide group that is the central one of a
three-site DB. In this case we conclude that peptide group can
provide the driving force amplitude to noise intensity ratio
$A/D$ up to $10$. For this purpose the DB should be favorably
situated relative the reaction coordinate, i.e., not far and at
favorable angle (see (\ref{eq8})). In the second case the field
is produced by three peptide groups of a five-site DB. In this case
we can attain $A/D$ up to $20\div30$.

The efficiency of the RPV as a dynamical contribution into enzyme
catalytic efficiency is evaluated by us within the framework of
the existing theories for the thermally activated escape in the
presence of the driving force \cite{Sme99}, \cite{Leh000}. It is
interesting to note that revealing the role of driving at
activated escape in biological systems is considered by the
authors of the paper \cite{Dyk01} as "a fundamentally important and most
challenging open scientific problem". We compile the
results of the theory in a simple phenomenological formula
(\ref{eq14}) that yields the required retardation of the
 exponential growth of the prefactor with the increase of $A/D$
 (see Fig.2) in accordance with the
results of \cite{Leh000}. Moreover  in this formula the phenomenon of
the so called "resonant activation" (see Fig.3), i.e., the
existence of an optimal frequency for reaction acceleration
manifests itself. The notion of resonant activation is
traditionally used for phenomena with the potential modulation by
random fluctuations \cite{Rei95}. However recent numerical
simulations of the Langevin equation \cite{Zol04} demonstrate
that this phenomenon takes place also for deterministic driving
and our result reproduces this computational finding. It should
 be stressed once more that the maximum of the frequency curves at
$\Omega\approx 0.2\div0.3$
 in Fig.3 is not related to the phenomenon of stochastic resonance
(defined by the requirement $\Omega=\pi \Gamma_K$
\cite{Gam98}) since the value of $\Gamma_K$ in our case is
 negligibly small ($\Gamma_K\approx 10^{-22}$).

With the help of the formula (\ref{eq14}) we conclude that the
RPV produced by a three-site DB can provide reaction acceleration
up to $10^3\div10^4$ (see Fig.2 at $A/D\approx10$ and
$\Omega\approx0.1\div0.3$). If we assume that the RPV is produced
by a five-site DB we can get $A/D$ up to $\approx20\div30$. In
this case we can obtain enormous reaction acceleration up to
$10^6\div10^8$ (see Fig.2 at $A/D\approx 20\div30$ and
$\Omega\approx0.1\div0.3$). Thus the present model has potential
capacity to explain significant part of the total effect of enzyme
efficiency. Moreover it gives an answer to the question why the
frequency range $\approx 100\div200\  cm^{-1}$ of the RPV was
chosen by the evolution of enzymes as an optimal one. This range
(dimensionless $\Omega \approx 0.1\div0.3$) turns out to be
favorite because it matches the requirement of resonant
activation where driving is most efficient for reaction
acceleration (see Fig.3).

Also the formula (\ref{eq14}) is shown to be compatible with
experimental data for solvent viscosity effect on enzymatic
reaction rates at a physically understandable assumption that the
amplitude of oscillations in the DB decreases with the increase
of solvent viscosity by the relationship (\ref{eq22}). The latter
actually means that with the increase of the number of cosolvent
molecules in the vicinity of the oscillating peptide group of
the DB the amplitude of the
oscillations is decreased because cosolvent molecules create
 steric hindrance for oscillations. For realization of this relationship a
separate model should be constructed. The latter seems to be a
feasible task and is planned for future work. Thus in our opinion the
present approach sheds new light on the origin of the dependence
of enzymatic reaction rates on solvent viscosity.

What kind of experiments can be interpreted with the help of the
present model? In its moderate form (reaction acceleration up to
$10^3\div10^4$ by a three-site DB) it can be directly applied to
explaining the residual catalytic activity of mutant enzymes. It
is known that even at substitution of all catalytically active
groups in the enzyme active site by inactive ones by methods of
gene engineering (when one should expect no lowering of the
potential barrier) the mutants still exhibit residual catalytic
activity accurately $10^3\div10^4$ \cite{Car88}, \cite{Bra99}.
Also the known phenomenon of catalytic antibodies can be treated.
These proteins are actually not enzymes. They have no active site
and no catalytically active groups at all \cite{Bra99}. Still they
also exhibit catalytic activity accurately $10^3\div10^4$
\cite{Bra99}. The explanation of chemical enzymology is that they
nevertheless bind the transition state of substrate molecules
preferentially thus stabilizing it \cite{Bra99}. The present
model suggests another explanation, namely, being proteins they
contain secondary structure in which DBs producing the RPV can be
excited at substrate binding. Another case if the extreme
estimates of the present model (reaction acceleration up to
$10^6\div10^8$ by a five-site DB) were necessitated and called
for by difficulties of traditional chemical enzymology. Such
enormous acceleration constitutes significant part of total
catalytic efficiency at enzyme action. Then the present model
would be applicable to interpreting much wider range of kinetic
data. We conclude that the present model for the dynamical
contribution into enzyme catalytic efficiency is able to describe
various situations from partial reaction acceleration up to major
part of the total effect and besides suggests a route to
revealing the origin of solvent viscosity effect on rate
constants. Whether or not nature makes use of such mechanism
remains the subject of further investigations.\\

Acknowledgements. The author is grateful to Prof. V.D. Fedotov,
 R.Kh. Kurbanov for helpful discussions. The work was supported
 by the grant from RFBR.

\newpage
\begin{figure}
\begin{center}
\includegraphics* [width=\textwidth]{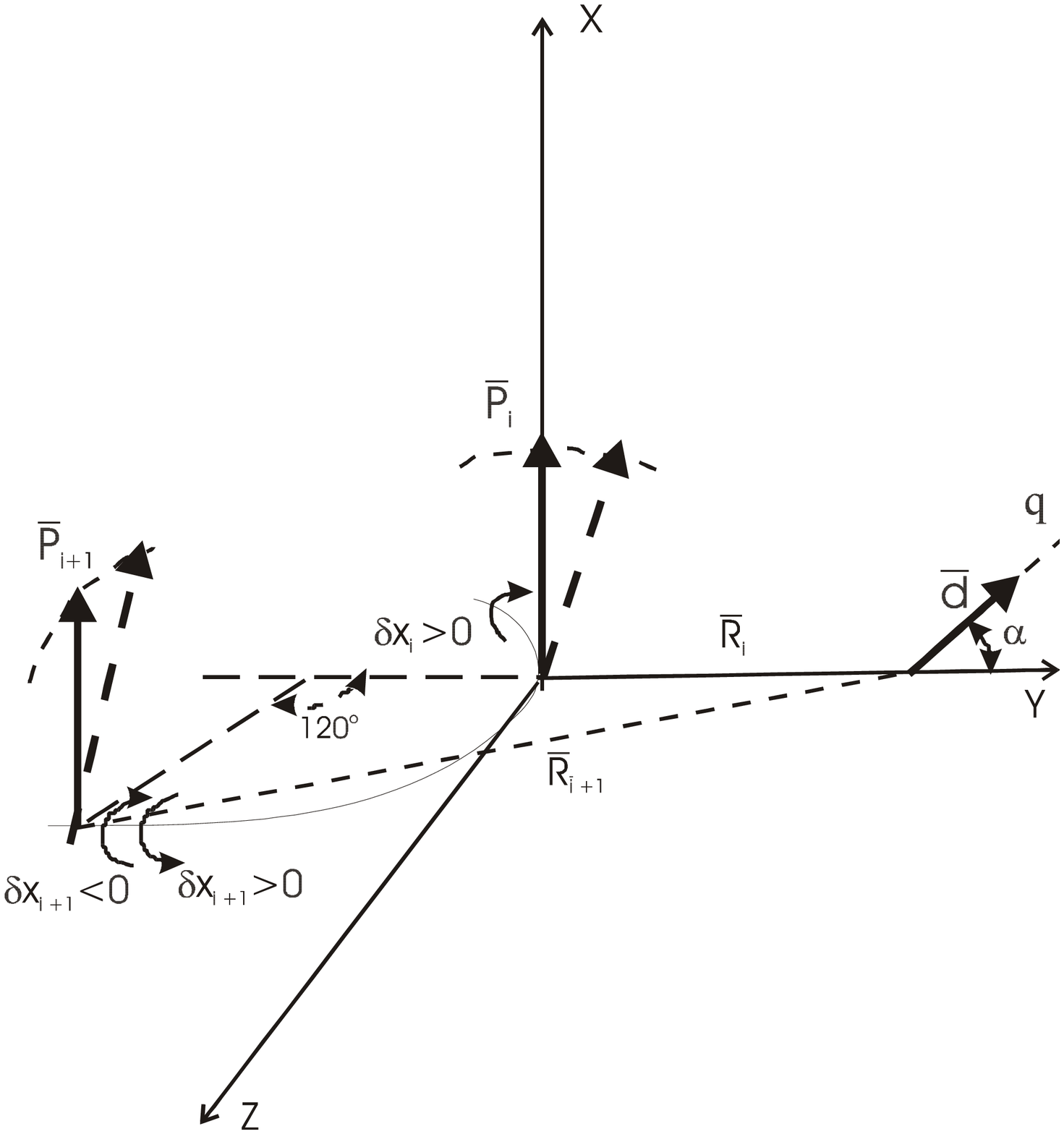}
\end{center}
\caption{Schematic representation of a part of an $\alpha$-helix
and a reaction coordinate $q$ at a distance $\bar R_i$ from it.
Here $\bar p_i$ is the dipole moment of the $i$-th peptide group
and $\bar d$ is the dipole moment of the reaction coordinate.}
\label{Fig.1}
\end{figure}

\clearpage
\begin{figure}
\begin{center}
\includegraphics* [width=\textwidth] {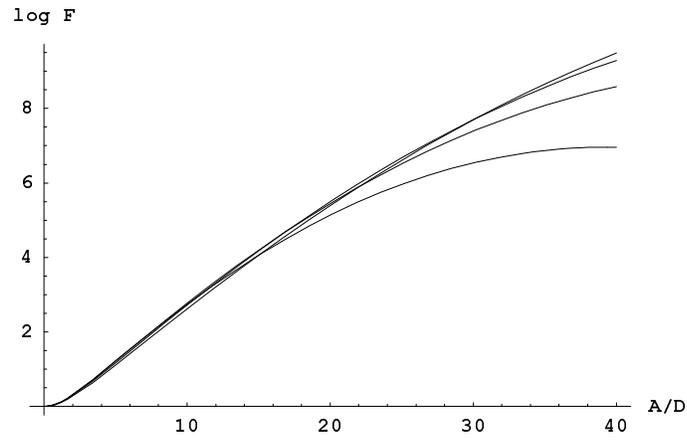}
\end{center}
\caption{The dependence of the reaction acceleration factor on
the driving force amplitude to noise intensity ratio obtained from
 (\ref{eq13}). The values
of the frequencies $\Omega$ of the driving force from the down
line to the upper one respectively are:
 0.15; 0.2; 0.25; 0.3. The value of the noise
 intensity is $D=3.3\cdot10^{-3}$.} \label{Fig.2}
\end{figure}

\clearpage
\begin{figure}
\begin{center}
\includegraphics* [width=\textwidth] {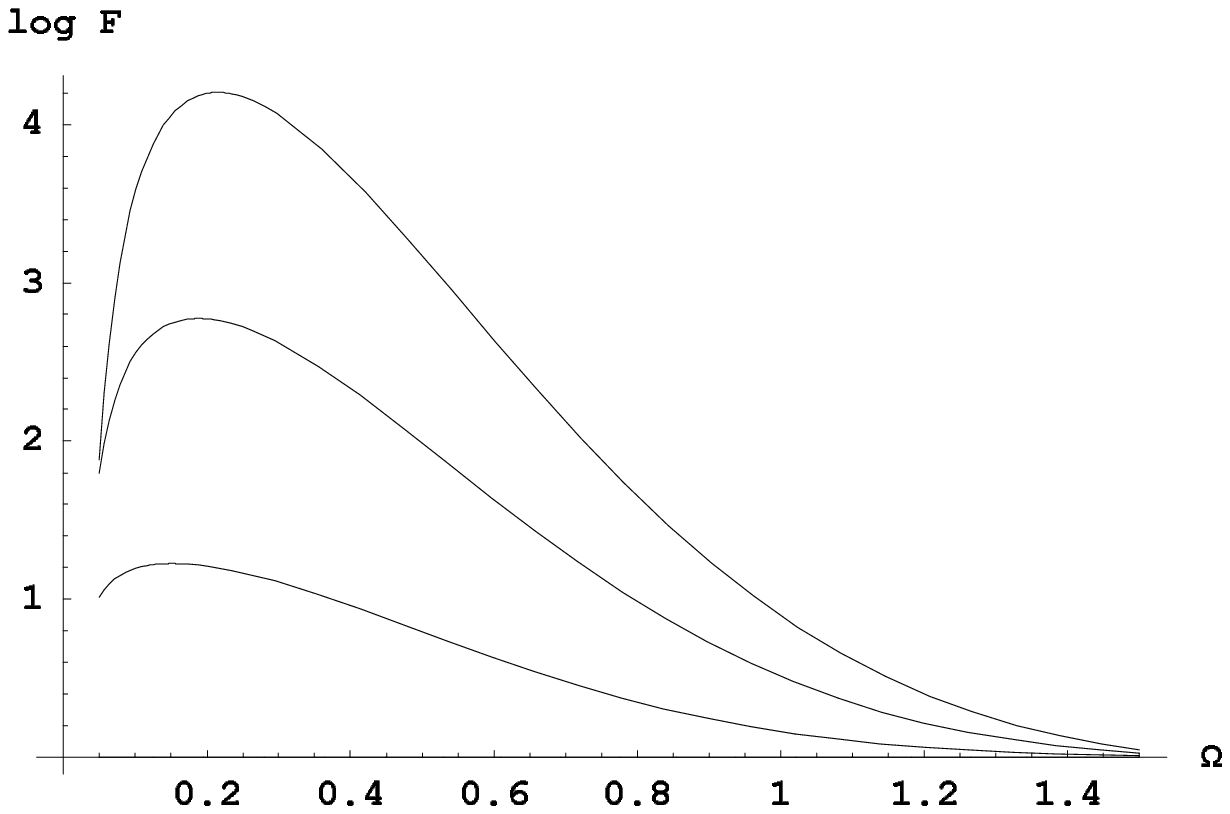}
\end{center}
\caption{The dependence of the reaction acceleration factor on
the frequency of the driving force obtained from (\ref{eq13}).
The values of the driving
amplitude of the periodic force to noise intensity ratio from the
down line to the upper one respectively are:
5; 10; 15.  The value of the noise
intensity is $D=3.3\cdot10^{-3}$.}
\label{Fig.3}
\end{figure}


\newpage
\begin{thebibliography}{00}
\bibitem{Kar79}
M. Karplus, J.A. McCammon, Protein structural fluctuations during
a period of 100 ps, Nature (London) 277 (1979) 578-578.
\bibitem{Wel82}
G.R. Welch, B. Somogyi, S. Damjanovich, The role of protein
fluctuations in enzyme action, Prog. Biophys., Mol.
Biol. 39 (1983) 109-146.
\bibitem{Kar83}
M. Kurplus, J.A. McCammon, Dynamics of Proteins: Elements and
Function, Annu.Rev.Biochem. 52 (1983) 263-300.
\bibitem{War84}
A. Warshel, Dynamics of enzymatic reactions,
Proc.Natl.Acad.Sci.USA 81 (1984) 444-448.
\bibitem{Flu85}
C.R. Welsh, ed., The fluctuating enzyme, Wiley, N.Y., 1986.
\bibitem{Ner97}
E. Neria, M. Karplus, Molecular dynamics of an enzyme reaction:
proton transfer in TIM, Chem.Phys.Lett. 267 (1997) 23-30.
\bibitem{Alp01}
K.O. Alper, M. Singla, J.L. Stone, C.K. Bagdassarian, Correlated
conformational fluctuations during enzymatic catalysis:
Implications for catalytic rate enhancement, Prot.Sci. 10 (2001)
1319-1330.
\bibitem{Agr05}
P.K. Agarwal, Role of protein dynamics in reaction rate
enhancement by enzymes, J.Am.Chem.Soc. 127 (2005) 15248-15256.
\bibitem{Fer99}
A. Fersht, Structure and mechanism in protein science: a guide to enzyme
catalysis and protein folding, Freeman, N.Y., 1999.
\bibitem{War97}
A. Warshel, Computer modeling of chemical reactions in enzymes and
solutions, Wiley, N.Y., 1997.
\bibitem{War98}
A. Warshel, Electrostatic origin of the catalytic power of
enzymes and the role of preorganized active sites, J.Biol.Chem.
273 (1998) 27035-27038.
\bibitem{Cle98}
W.W. Cleland, P.A. Frey, J.A. Gerlt, The low barrier hydrogen bond
in enzymatic catalysis, J.Biol.Chem. 273 (1998) 25529-25532.
\bibitem{Ant01}
D. Antoniou, S.D. Schwartz, Internal enzyme motions as a source
of catalytic activity: Rate-promoting vibrations and hydrogen
tunneling, J.Phys.Chem. B105 (2001) 5553-5558.
\bibitem{Sut02}
M.J. Sutcliffe, N.S. Scrutton, A new conceptual framework for
enzyme catalysis. Hydrogen tunneling coupled to enzyme dynamics
in flavoprotein and quinoprotein enzymes, Eur.J.Biochem. 269
(2002) 3096-3102.
\bibitem{Ant02}
D. Antoniou, S. Caratzoulas, C. Kalyanarman, J.S. Mincer, S.D.
Schwartz, Barrier passage and protein dynamics in enzymatically
catalyzed reactions, Eur.J.Biochem. 269 (2002) 3103-3112.
\bibitem{Kna02}
M.J. Knapp, J.P. Klinman, Enviromentally coupled hydrogen
tunneling. Linking catalysis to dynamics, Eur.J.Biochem. 269
(2002) 3113-3121.
\bibitem{Min03}
J.S. Mincer, S.D. Schwartz,  Protein promoting vibrations in
enzyme catalysis. A conserved evolutionary motif, J.Proteome Res.
2 (2003) 437-439.
\bibitem{Min04}
J.S. Mincer, S.D. Schwartz,  A computational method to identify
residues important in creating a protein promoting vibration in
enzymes, J.Phys.Chem. B107 (2003) 366-371.
\bibitem{Yed95}
S. Yedgar, C. Tetreau, B. Gavish, D. Lavalette, Viscosity
dependence of $O_2$ escape from respiratory proteins as a
function of cosolvent molecular weight, Biophys.J. 68 (1995)
665-670.
\bibitem{Dan95}
N. More, R.M. Daniel,  H.H. Petach, The effect of low
temperatures on enzyme activity, Biochem. J. 305 (1995) 17-20.
\bibitem{Dan98}
R.M. Daniel, J.C. Smith, M. Ferrand, S. Hery, R. Dunn, J.L.
Finney, Enzyme activity below the dynamical transition at 220 K,
Biophys.J. 75 (1998) 2504-2507.
\bibitem{Bra00}
J.M. Bragger, R.V. Dunn, R.M. Daniel, Enzyme activity down to
$-100^{\circ}$, BBA 1480 (2000) 278-282.
\bibitem{Dun00}
R.V. Dunn, V. Reat, J. Finney et al., Enzyme activity and dynamics:
xylanase activity in the absence of fast anharmonic dynamics,
Biochem. J. 346 (2000) 355-358.
\bibitem{Ibe89}
E.T. Iben, D. Braunstein, W. Doster et al., Glassy behavior of a
protein, Phys.Rev.Lett. 62 (1989) 1916-1919.
\bibitem{Kov99}
V.A. Kovarskii, Quantum processes in biological molecules. Enzyme
catalysis, Usp.Phys.Nauk, 169 (1999) 889-908.
\bibitem{Sit03}
A.E. Sitnitsky, Discrete breathers in protein secondary
structure, in: "Soft condensed matter. New research."  Ed. K.I.
Dillon, Nova Science Publishers Inc., NY,  2006;  Los Alamos
preprint datebase, arXiv:cond-mat/0306135.
\bibitem{Siev88}
A.J. Sievers, S. Takeno, Intrinsic localized modes in anharmonic
cristals, Phys.Rev.Lett. 61 (1988) 970-974.
\bibitem{Mac94}
R.S. MacKay, S. Aubry, Proof of existence of breathers for
time-reversible or harmonic networks of weakly coupled
oscillators, Nonlinearity  7 (1994) 1623-1629.
\bibitem{Fla98}
S. Flach, C.R. Willis, Discrete breathers, Phys.Reports 295
(1998) 181-264.
\bibitem{Xie02}
A. Xie, A. van der Meer, R.H. Austin, Excited-state lifetimes of
far-infrared collective modes in proteins, Phys.Rev.Lett. 88
(2002) 018102-1-018102-4.
\bibitem{Can80}
C.R. Cantor, P.R. Schimmel, Biophysical Chemistry, Freeman, San
Francisco, 1980, part 1.
\bibitem{Dyk01}
 M.I. Dykman, B. Golding, L.I. McCann et al., Activated escape of
periodically driven systems, Chaos 11 (2001) 587-598.
\bibitem{Han90}
P. H\"anggi, P. Talkner, M. Borkovec, Fifty years after Kramers'
equation: reaction rate theory, Rev.Mod.Phys. 62 (1990) 251-341.
\bibitem{Jun93}
P. Jung, Periodically driven stochastic systems, Phys.Reports 234
(1993) 175-295.
\bibitem{Gam98}
L. Gammaitoni, P. H\"anggi, P. Jung, F. Marchesoni,  Stochastic
resonance, Rev.Mod.Phys. 70 (1998) 223-287.
\bibitem{Leh00}
J. Lehmann, P. Reimann, P. H\"anggi, Surmounting oscillating
barriers, Phys. Rev. Lett. 84 (2000) 84-87.
\bibitem{Leh000}
J. Lehmann, P. Reimann, P. H\"anggi, Surmounting oscillating
barriers: Path-integral approach for weak noise, Phys.Rev. E62
(2000) 6282-6294.
\bibitem{Leh03}
J. Lehmann, P. Reimann, P. H\"anggi, Activated escape over
oscillating barriers: The case of many dimensions,
Phys.stat.Sol.(b) 237 (2003) 53-64.
\bibitem{Sme99}
V.N. Smelyanskiy, M.I. Dykman, B. Golding, Time oscillations of
escape rates in periodically driven systems, Phys.Rev.Lett. 82
(1999) 3193-3197.
\bibitem{Sme99a}
V.N. Smelyanskiy, M.I. Dykman, H. Rabitz et al., Nucliation in periodically
driven electrochemical systems, J.Chem.Phys. 110 (1999) 11488-11504.
\bibitem{Mai01}
R.S. Maier, D.L. Stein, Noise-activated escape from a sloshing potential well,
Phys.Rev.Lett. 86 (2001) 3942-3946.
\bibitem{Tal99}
P. Talkner, Stochastic resonance in the semiadiabatic limit, New J.Phys. 1
(1999) 4.1-4.25.
\bibitem{Tal04}
P. Talkner, J.Luczka, Rate description of Fokker-planck processes with time
dependent parameters, Phys.Rev.E 69 (2004) 046109.
\bibitem{Ryv04}
D. Ryvkine, M.I. Dykman, B. Golding, Scaling and crossovers in
activated escape near a bifurcation point, Phys. Rev. E 69 (2004)
061102.
\bibitem{Dyk05}
M.I. Dykman, D. Ryvkine, Activated escape of periodically
modulated systems, 94 (2005) 070602.
\bibitem{Dyk051}
D. Ryvkine, M.I. Dykman, Noise-induced escape of periodically modulated systems:
From weak to strong modulation, Phys.Rev E72 (2005) 011110.
\bibitem{Rei95}
P. Reimann, Thermally driven escape with fluctuating potentials: A new type
of resonant activation, Phys.Rev.Lett. 74 (1995) 4576-4579.
\bibitem{Bra99}
C. Branden, J. Tooze, Introduction to protein structure, 2-nd
ed., Garland Publishing Inc, N.Y., 1999.
\bibitem{Car88}
P. Carter, J. Wells, Dissecting the catalytic triad of a serine
protease, Nature (London) 332 (1988) 564-568.
\bibitem{Zol04}
Y. Zolotaryuk, V.N. Ermakov and P.L. Christiansen, Resonant enhancement of
the jump rate in a double-well potential, J.Phys. A: Math.Gen. 37 (2004)
6043-6051.


\end{thebibliography}
\end{document}